\documentclass[journal,comsoc]{IEEEtran}

\usepackage[T1]{fontenc}

\usepackage{cite} 
\ifCLASSINFOpdf
  \usepackage[pdftex]{graphicx}
\else
\fi

\usepackage{amsmath}
\usepackage[cmintegrals]{newtxmath}

\usepackage{listings}
\usepackage[belowskip=-5pt,aboveskip=0pt]{caption}
\begin{document}

\title{IoT-based Smart Home Device Monitor Using Private Blockchain Technology and Localization}

\author{Marc Jayson Baucas,~\IEEEmembership{Student Member,~IEEE,} Stephen Andrew Gadsden,~\IEEEmembership{Senior Member,~IEEE,}\\ Petros Spachos,~\IEEEmembership{Senior Member,~IEEE}
\thanks{Marc Jayson Baucas, Stephen Andrew Gadsden and Petros Spachos are with the School of Engineering, University of Guelph, Canada, N1G 2W1.}}

\maketitle

\begin{abstract}
Internet of Things (IoT)-based smart home applications are rising in popularity. However, this trend attracts malicious activity, which causes cost-efficient security to be in high demand. This paper proposes a low-end design that reinforces the security of a home network. It uses private blockchain technology and localization via RSSI-based trilateration.  We investigated the benefits of private blockchains over their public counterpart, and we improve the precision of the localization algorithm by testing it against different wireless technologies. The results conclude that using a private blockchain with a WiFi-based communication system produces the most efficient iteration of the proposed design. 

\end{abstract}


\section{Introduction}  
Several smart home services relying on using Internet of Things~(IoT) devices. Implemented services such as remove surveillance and personal data storage create a network of devices that provide control over the different smart systems within the house~\cite{collotta}. However, many home networks fall victim to attackers that infiltrate the server and tamper with it. Also, aside from the standard data router, these networks lack a dedicated administrator to regulate their connected devices. To address these issues, a secure network administrator is needed~\cite{li}. By doing so, it can remain safe and monitored properly against security breaches.

This work focuses on creating a low-end solution to manage the devices attempting to access the network. We use the built-in trust-based protocol of private blockchains to create a tamper-proof and secure administrator. Also, due to its architecture, we can automate the network for more efficient management. However, using this technology limits it to track what goes in and out of the network. Therefore, it cannot pinpoint the source of malicious activity in terms of location~\cite{anthi}. To address this, we propose the use of a location-based filter via distance trilateration using the Received Signal Strength Indicator (RSSI). Trilateration-based indoor localization allows the network to pinpoint any device that is within the working radius of the network~\cite{sahin}. To further improve the precision of the collected RSSI values, we chose to incorporate Kalman filtering to manage the raw data~\cite{kalman-iot}.


\section{Background and Motivation}
\subsection{Smart home security based on IoT}
Smart homes are an implementation of IoT systems within a home network~\cite{li}. It is a wireless home network that provides services and applications developed for improved health, comfort, and user safety~\cite{collotta}. However, this amount of data within it is vulnerable to intrusive attacks in terms of regulating device access~\cite{anthi}. Therefore, by adding a secure medium that monitors the devices attempting to access the network, data can remain private.

Smart homes have become a popular option for modern households. However, due to this increase in interest, security attacks towards remote monitoring and control are introduced~\cite{komninos}. An example is client impersonation. Client impersonation is when a malicious user can infiltrate and compromise the privacy of a homeowner. As a result, they gain complete control over their remote services. Our paper suggests solutions to fortify the security of smart home networks against such attacks. 

In~\cite{anthi}, they proposed a three-layer intrusion detection system (IDS) that uses a supervised machine learning approach to detect cyber-attacks on IoT networks. As for ~\cite{arif}, they take a different approach by suggesting blockchains to regulate their proposed IDS. Although these designs, on their own, can detect an attack without knowing where the attack is coming from, it is not possible to implement further prevention. Therefore, our proposed framework adapts to these designs using indoor localization via RSSI trilateration via Kalman filtering to add the ability to trace the sources of the attack.

\subsection{Blockchain}
Blockchains are data structures composed of a chain of blocks that are cryptographically linked~\cite{beekeeper}. It can be a historical tamper-proof ledger for data and transaction management~\cite{b-meets-iot}. This design is more useful in IoT architectures due to its ability to preserve the integrity of the network's data~\cite{b-survey}~\cite{li-edge}. Blockchains can also automate their processing with the incorporation of smart contracts. A smart contract is a term-based transaction protocol with a defined function and an assigned activation agreement~\cite{b-enabled-smart}. Also, blockchains use a consensus process called "mining" to determine the actions and changes carried out within their system. We plan to use this process to filter and give access to connecting devices in the network.

There are two models of authentication in blockchains: public and private~\cite{block-iot}. A public blockchain relies on a proof-of-work system~\cite{bitcoin}. This system requires incoming users to solve a provided algorithm to grant them voting rights to be in the mining process. Executing this complex algorithm requires high processing power, which is not ideal for IoT devices due to their processor constraints~\cite{block-iot}. On the other hand, a private blockchain uses a built-in trust-based access layer to authorize users~\cite{b-meets-iot}. This feature eliminates the need for a proof-of-work system~\cite{baucas}. However, as the ledger grows, latency becomes an issue. Therefore, there is a visible trade between the two blockchains in terms of latency and resource management~\cite{b-survey}. Since our proposed network is for a smaller group of users and low-end devices, a private blockchain is better. However, we integrate RSSI-based indoor localization via trilateration to further improve the detection system of the network. 

\subsection{RSSI-based Indoor Localization via Trilateration}
Wireless indoor localization is a concept of determining the location of a point of interest (POI) within a controlled environment~\cite{sahin}. It takes advantage of the signal characteristics that devices provide to estimate the location of each device~\cite{shu}. One of the most commonly used signal features is the RSSI value. This value is a metric for the power of a wireless signal. 



Trilateration is a method that is for localization. It uses the distances between a point of interest and three known points, also known as anchor nodes, to solve for the position of this point on a 2-dimensional plane~\cite{sadowski}. 




However, wireless signals are still subject to noise as the number of potential sources of interference within an area increases~\cite{sadowski}. Therefore, to ensure that the localization of these points of interest is accurate, Kalman filtering is used.

A Kalman filter (KF) is an estimating algorithm commonly used for linear systems~\cite{kalman-iot,gad}. Some of its variations like the Extended Kalman or the Unscented Kalman Filter usually veers toward a non-linear system~\cite{iot-loc}. A standard KF was selected to model this algorithm due to the linearity of distance and localization via RSSI. This selection also benefits the device since it requires less processor power to carry out the algorithm. Since we intend the frequent usage of the trilateration algorithm within the framework, a less complex equation is better in terms of scalability and long-term use.

A standard KF uses the state estimate ($\widehat{x}$) and the state estimate covariance ($P$). The state estimate is the predicted future value of the system. Meanwhile, the state estimate covariance is the approximated accuracy of the state estimate. The filtering process of a KF is composed of two steps; the predicting and updating step.

The prediction step estimates the next state estimate and covariance values of the system in its current time index,~$k$. This process is: 

\begin{equation}\begin{split}
  \widehat{x}_{k+1|k} &= A_{k}\widehat{x}_{k} + B_{k}u_{k}\\
P_{k+1|k} &= A_{k}P_{k}A_{k}^{T} + Q_{k}
\end{split}\label{x-p}\end{equation}
using the system model (A), measurement model (B), control input (u) and noise covariance (Q). 

The updating step is carried out by first solving for the Kalman gain (K) as:

\begin{equation}
  K_{k+1} = P_{k+1|k}C_{k+1}^{T}(C_{k+1}P_{k+1|k}C_{k+1}^{T} + R_{k+1})^{-1}\\
  \label{kgain}
\end{equation}
This equation introduces the measurement sensitivity (C) and measurement error covariance (R). Next, the resulting values are the indexed final state estimates and covariance values of the system. This finalization process as it incorporates the measured input (z) is:

\begin{equation}
  \begin{split}
    \widehat{x}_{k+1|k+1} &= \widehat{x}_{k+1|k} + K_{k+1}(z_{k+1} - H_{k+1}\widehat{x}_{k+1|k})\\
    P_{k+1|k+1} &= (1-K_{k+1}H_{k+1})P_{k+1|k}(1-K_{k+1}H_{k+1})^{T} \\ 
           & + K_{k+1}R_{k+1}K_{k+1}^{T}
  \end{split}
  \label{xpf}
\end{equation} 

This process is recursive to produce an estimated representation of the filtered data. 

\section{Methodology}\label{meth}
\subsection{Overview}
The proposed design focuses on a low-end access level implementation for smart home networks. This access layer is composed of two components to manage the devices within the network. The first component uses a private blockchain to determine any unrecognized devices attempting to connect to the network.  It can create a list of all the trusted devices based on the stored information on the server. As a result, the network can manage which connected devices can access and contribute to its stored data. However, this type of protection towards unauthorized access may not be enough to keep the network protected.

The second component will use localization to gain more information about the source of the attack. It is to determine the general location of the device. This filter will allow the ledger to keep track of the projected positions of the devices. It uses anchors that pinpoint the location of connecting devices via trilateration. To further increase the accuracy of the process, we decided to incorporate a simple Kalman filter. By integrating estimation theory into the design, the localization of devices can allow more accurate monitored node positions around the sensing network. 

The resulting network model starts with the main network router. Around it is the anchors that include a copy of the blockchain implemented to a home network. This cluster encompasses the unit that will regulate and manage the device authorization as the administrative head. Around each anchor are the smart appliances and devices that attempt to access the different home network services. With these components, this paper proposes low-end access and device managing system for smaller sensing networks.
 
\subsection{System Components}
The proposed design uses Raspberry Pi 3 Model B's as its main centre of operations for data filtering. It serves as the anchors that create the perimeter around the network. We selected this device due to its modularity and rapid prototyping. Also, the Pis were to represent the devices and appliances within a smart home. Some smart devices have the bare minimum to have a compact and optimized design. Therefore, to simulate these low-cost devices, we chose to use a Pi. 

We programmed each Pi with the same Raspbian-Jesse OS image. Also, we used Python 3.7 in incorporating software components needed in the design. These main parts include the blockchain and the Kalman filter. The blockchain component is programmed using python initialized upon executing the Pi. Each blockchain contains the same configuration and list of trusted devices. Within each anchor are its unique identifier and geographical position on the network. We embedded the Kalman filter within the blockchain class as a smart contract triggered by a data transaction. The RSSI values will is for calculating the location of each device.

\section{Results}\label{result}

\subsection{Testbed}

\begin{figure}[t!]
  \centering
  \includegraphics[width=\linewidth]{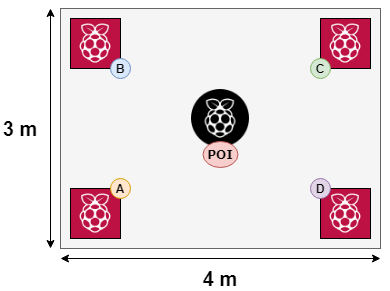}
  \caption{Room setup for access layer testbed showing anchor locations and point of interest.}
  \label{roomsetup} 
\end{figure}

We tested each aspect of the access layer separately to optimize the proposed design. The testbed has four anchors arbitrarily placed around a room shown in Fig.~\ref{roomsetup}. We recognize that more Pis can be added to the testbed to model more complex infrastructures. However, we chose to test our framework on simpler setups to ensure that the design foundation is plausible. 

The first test compares public and private blockchains to determine which is more efficient at filtering devices. The metric used is the memory, processor usage, and execution time. The test is designed with a device requesting access to the server through an anchor. Each anchor consults the blockchain with which users are allowed to connect. For the public blockchain, it processes the request through proof of work. The algorithm used is a complex comparator using the nonce of the block. It will require the anchor to increment the nonce until it reaches the desired value with an arbitrary number of leading zero bits. Listing~\ref{proofOfWork} shows the algorithm used. As for private blockchains, the anchor will attempt to consult through the trusted ledger. We preloaded the blockchains with the identification of devices that are authorized to access the network. Obtaining the physical memory usage was through a python library called memory-profiler. Meanwhile, we used the built-in CPU usage Monitor of the Pi to get the processor usage. As for execution time, it is the elapsed time between a device sends a request and the anchor responds.

\begin{lstlisting}[float=t!, language=Python, breaklines, frame=single, captionpos=b, caption={Proof of work code abstraction.}, label={proofOfWork}, belowskip=-2ex, basicstyle=\fontsize{8}{9}\selectfont\ttfamily]
def proof_of_work(block, n):
  block.nonce = 0
  block_hash = sha256(block) 
  while not block_hash[:n] = '0' * n:
    block.nonce += 1
    block_hash = sha256(block)
  return hash
\end{lstlisting}

We carried out the second test by focusing on the localization via a trilateration filter. The created $4\times3$~m workspace simulates a room as the scope of the access layer, as shown in Fig.~\ref{roomsetup}. We will decide between BLE, WiFi, or ZigBee to maximize the capabilities of the Kalman filter. These capabilities are on their accuracy in representing the distance between the anchor and the device. We set up the BLE transmissions using the built-in Bluetooth drivers within the Pis. As for WiFi, we used a WiFi-based network scanning python library called Rssi. Lastly, for ZigBee, a Digi XBee S2C module is installed with the python-XBee library. We calculated the distance with parameters calibrated based on the hardware specifications of the wireless components used with the Pis. All technologies used a 2.00 Path Loss Factor. For the system loss constant, BLE uses -56, WiFi uses -45, and XBee uses 18. To compare the precision of the different wireless technologies, we calculated the Root Mean Squared Error (RMSE) as:
 
\begin{equation}
  RMSE = \sqrt{\frac{\sum_{i=1}^{n} (P_{i} - O_{i})^{2}}{n}}
  \label{rmse}
\end{equation}
where $P$ represents the predicted value and $O$ is the observed under $n$ time samples. RMSE highlights the impact of the Kalman filter in maintaining the consistency of the distance calculations.

\subsection{Blockchain evaluation}
The blockchain comparison measures the memory usage, CPU usage, and elapsed time between the anchor consulting the blockchain and the moment it responds. We conducted the test for 20 iterations of the anchor accessing each blockchain type. Table~\ref{test-bc} shows the results of testing for each blockchain technology.

\begin{table}[t!]
  \centering
  \small
  \begin{tabular}{|p{0.2\columnwidth}|p{0.2\columnwidth}|p{0.2\columnwidth}|p{0.2\columnwidth}|}
    \hline
     \textbf{Blockchain Type} & \textbf{Memory Usage (MB)} & \textbf{CPU Usage (\%)} &\textbf{Execution Time (s)} \\ \hline
     Private  & 12.20  & 11         & 0.00021   \\ \hline
     Public & 12.37    & 25       & 25.89   \\ \hline
  \end{tabular}
  \caption{Memory usage and execution time results of the blockchain tests.}
  \label{test-bc}

\end{table}

In terms of memory usage, both types of blockchains use relatively the same amount of memory. These values show that running these blockchains on the anchor requires the same amount of allocated memory to access. However, accessing the private blockchain takes 14\% less CPU usage than its public counterpart. This difference shows how private blockchains take fewer resources to run on the Raspberry Pi. Also, a transaction between the anchor and the public blockchain takes 25 seconds. Meanwhile, the private blockchain is 0.21 milliseconds. This discrepancy in the execution time could be the proof of work required for the public blockchain. Due to the expected volume of transactions between the anchor and the blockchain, a lower resource demand and execution time is better. Therefore, choosing to use private blockchains in the proposed design proves to be better.

\subsection{Localization filter evaluation}
\begin{figure}[t!]
  \centering
  \includegraphics[width=\columnwidth]{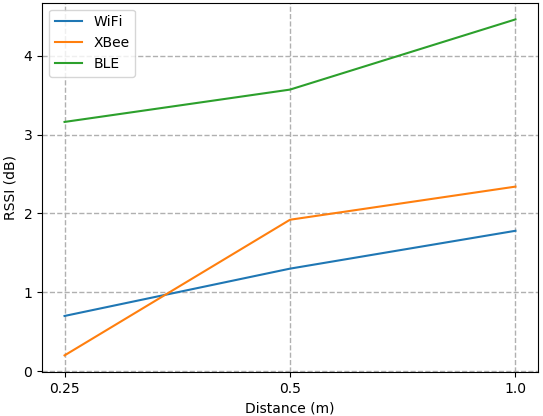}
  \caption{RMSE of RSSI values from WiFi, XBee, and BLE.}
  \label{rssi_rmse}
  \vspace{-2ex}
\end{figure}

We tested the proposed design based on its precision in estimating the distance between an anchor and a point of interest. The testbed has the device at varying distances from the anchor on a flat plane. The values used in the tests were; 0.25, 0.50 and 1 m. The RSSI RMSE results of each wireless technology are shown in Fig.~\ref{rssi_rmse} as collected over 100 samples.

Based on the obtained metrics, the consistency of the RSSI decreased as the distance between the device and the anchor increased. As a result, when the actual distance increases, the preciseness of the theoretical value decreases. Comparing the three technologies, XBee had the best results in terms of distances closer to the device. However, it could not maintain a reasonable trend in RSSI consistency as the actual distance increased. For BLE, it showed the most inconsistent RSSI values among the three. This observation shows how it is the least ideal in terms of precision. Meanwhile, WiFi yielded the most consistent RSSI reported among the three technologies. Overall, WiFi had the best results in minimizing the effects of the actual distance to the precision of the calculated value.

\section{Conclusion}\label{conc}
The proposed design is a combination of private blockchain technology and localization via RSSI-based trilateration. It aims to create an access layer that can increase the security of home networks. We conducted two tests to check which technologies were more optimal. The first was to compare the presented types of blockchains for memory usage, CPU usage, and execution times. The results show that private blockchains proved to be better for the design. The second test checks which communication medium among BLE, WiFi, and ZigBee yielded the most precise RSSI generation. The results showed that WiFi stayed the most consistent. Therefore, the design is more optimized by integrating private blockchains over public and WiFi for RSSI measurements. Overall, the design has proven its plausibility as an access layer that reinforces security for smart home networks.

\bibliographystyle{IEEEbib}
\bibliography{IEEEabrv,rssiestbib}
\end{document}